\documentstyle[12pt]{article}

\large
\newcommand{\be}{\begin{eqnarray}}
\newcommand{\ee}{\end{eqnarray}}

\begin{document}
\setlength{\baselineskip}{21pt}
\pagestyle{empty}
\vfill
\eject
\begin{flushright}
SUNY-NTG-94-22
\end{flushright}

\vskip 1.25cm
\centerline{\bf QCD Instantons in Vacuum and Matter}
\vskip 1.25cm
\centerline{I. Zahed}
\vskip .2cm
\centerline{Department of Physics}
\centerline{SUNY, Stony Brook, New York 11794}
\vskip 1.25cm

\centerline{\bf Abstract}
\vskip 3mm

The singlet coupling to the topological charge density
in the instanton vacuum, causes the  instantons and antiinstantons to be
screened
over distances of the order of $1/2$ fm. Dilute instanton systems behave as a
free gas, while dense instanton systems behave as a plasma. The free gas
behaviour is favored by a density of 1 fm$^{-4}$.
Owing to the Higgs mechanism, the $\eta$' mass is heavy ($1100$ MeV).
The vacuum topological susceptibility is small ($0.07$ fm$^{-4}$) and
consistent with the QCD Ward identity. In the chiral limit, the singlet
screening vanishes, leading to a four dimensional plasma state with a
temperature given by the nonet decay constant $f$ (90 MeV).
The phase is Debye screened. In the presence of matter, the screening is
quenched, and a Kosterlitz-Thouless transition may occur signaling the
restoration of the $U_A(1)$ symmetry. We suggest to use the spatial
asymptotics of the static topological charge correlator together with the
topological susceptibility to probe the interplay between the $U_A(1)$
restoration and the chiral restoration in finite temperature QCD.
The relevance of these results to the bulk thermodynamics in the instanton
vacuum is discussed.

\vfill
\noindent
\begin{flushleft}
SUNY-NTG-94-22\\
April 1994
\end{flushleft}
\eject
\pagestyle{plain}

{\bf 1. Introduction}
\vskip .5cm

The instanton approach to the QCD ground state has received continuous
attention in the past few years
\cite{DIA1,SHU1,NVZ1,VS}. Overall, it provides a fair description
of the bulk vacuum properties and hadronic correlations.
Recent (quenched) lattice simulations \cite{NEG}
have suggested that the bulk structure
and hadronic correlations are mildly affected by cooling, an indication that
semiclassical physics may be operative in the long wavelength limit.

Extensive calculations using instantons have been carried out in the quenched
approximation (equivalently large $N_c$ limit), both analytically
\cite{DIA2,NVZ2} and numerically \cite{VS}. While the numerical calculations
were found to be consistent with the
cooled lattice simulations, these results suggest that the instantons in the
vacuum state are in a dilute gas approximation. To the annoying exception
of the $\eta$' mass, which comes out to be too heavy in the numerical instanton
calculations, the overall part of the
hadronic spectrum is reproduced by fitting the pion decay constant and the
fermion condensate. The model does not confine, yet it breaks spontaneously
chiral symmetry with the generation of a momentum dependent constituent mass.

To understand some of the features of the model, it suffices to recall
few thinghs. The pseudoscalar nonet follows from symmetry  provided that
the pion decay constant and the fermion condensate are reasonable, given
suitable current quark masses. In the dilute gas approximation, the vector and
isobar correlators are not affected by instantons at small distances. The
formers because of the self-duality character of the instanton configuration,
and the latters because of the flavor character of the induced instanton
interaction.
Interesting effects in the nucleon channel have been reported suggesting
diquark correlations at intermediate distances \cite{VS}.
Since the model lacks
confinement, the large distance behaviour of these correlators is
dominated by constituent quark thresholds.

A major unknown in instanton calculations resides in  the character of the
instanton-antiinstanton interaction. Earlier calculations based on
variational estimates with a sum ansatz in the quenched approximation,
have led to a reasonable description of the bulk structure \cite{DIA3}.
However, these estimates were found to
be quantitativaly sensitive to the short distance character of the interaction.
The latter was ansatz dependent. More elaborated numerical analyses, based
on stream-line configurations as the 'best' instanton-antiinstanton prototypes,
have shed question marks on the use of a semiclassical approximation
altogether \cite{V1}. However, these configurations hold only in the quenched
approximation, where the instanton-antiinstanton interaction is
dipole-like at large distances. They may not be important in the unquenched
approximation, where screening by charges in the fundamental representation
is in effect.

Semiclassical physics in QCD has been challenged by Greensite \cite{GREEN}
using a combination of arguments based on large $N_c$, factorization and the
master field equation. Greensite's objections to instantons (and generically to
semiclassical models based on statistical ensembles) may be evaded
by noting that the large $N_c$ limit does not commute with the thermodynamical
limit.

In this paper we would like to come back to the annoying problem in the
$\eta$' channel. Since instantons have been professed to cure the $U_A(1)$
problem \cite{HOOFT}, this discrepancy cannot be ignored. Also,
in the unquenched approximation, the instanton-antinstanton interaction
is not dipole-like at large distances, but screened over distance scales of
the order of $1/2$ fm. This screening is caused by the mixing between the
singlet and the topological density in the vacuum, and is at the origin of
the large $\eta$' mass, thus resolving the $U_A(1)$ problem. In fact this
construction was briefly presented in \cite{NVZ2} and more recently
discussed in the context of the strong $CP$ problem \cite{CP}.
We will also show that the resulting vacuum topological susceptibility is
consistent with the QCD Ward identity. In the unquenched approximation, dilute
systems of instantons behave as a free gas. Throughout, the short distance part
of the instanton-antiinstanton interaction will be assumed.
In the presence of matter the screening
decreases. Our approach suggests a dual description for the bulk pressure :
either as  a free meson gas with a heavy $\eta$', or as a free meson gas with
a light singlet plus a screened Coulomb gas of instantons and antiinstantons.
At temperatures for which the screening length becomes comparable to
the interparticle distance, a Kosterlitz-Thouless transition \cite{KT}
may occur. This transition may be related to the $U_A(1)$ restoration in QCD.
We suggest to use the large spatial asymptotics of the topological charge
correlator to account for the restoration of $U_A(1)$ in the QCD ground
state. We explicitly show that the bulk thermodynamics at low temperature
is meson dominated, and a rapid crossover in the topological susceptibity is
expected with the vanishing of the fermion condensate.
Throughout, the octet-singlet mixing will be ignored and the vacuum angle
will be set to zero.

\vskip 1cm
{\bf 2. Model}
\vskip .5cm

Few years ago, Nowak, Verbaarschot and I \cite{NVZ2} suggested
that in the long wavelength approximation
the salient features of the instanton model to the QCD ground state follows
from symmetries and anomalies. If we were to denote by
$n^{\pm}$ the local densities of instantons and antiinstantons
(thought about here as quasiparticles), then
the partition function in the presence of a pseudoscalar
source term $S^a$, reads \cite{NVZ2}

\be
Z [S] = \int dn^{\pm} \,\,\int d K \,\, e^{-W [K] - W_I [K, n] +
\int d^4x {\rm Tr}(S K)}
\label{1}
\ee
where $W[K]$ is the pseudoscalar effective action

\be
W[K] = \int d^4x \left( \frac 12 (\partial_{\mu} K^a)^2 -\frac 12
\frac{<\overline{\psi}\psi >}{f^2}\,\,{\rm Tr}(m K^2)\right)
\label{2}
\ee
and  $W_I [K, n]$ is the singlet coupling to the local topological charge
$(n^+-n^-)\sim E^a\cdot B^a$,

\be
W_I [K, n] =-i \int d^4x \frac{\sqrt{2N_F}}{f} (n^+ -n^-) \,\,K^0 (x)
\label{3}
\ee
We have defined the nonet by $K= \sum_{a=0}^8 K^a\lambda^a$, with the
normalization ${\rm Tr} (\lambda^a\lambda^b )=\delta^{ab}$. The mass matrix
$m={\rm diag}\, (m,m, m_s)\sim {\rm diag}\, (5,5,120)\,\,$ MeV.
The imaginary part follows from the fact that
(\ref{3}) is T-odd in Minkowski space. The singlet
coupling to the topological charge is
$\sqrt{2N_F}/f$, where $f\sim 90$ MeV is the bare nonet decay constant, and
$N_F$ the number of flavours. It vanishes in the large $N_c$ limit as it
should. The fermion condenstate
$<\overline\psi \psi>\sim - (255 \,\,{\rm MeV} )^3$, for an instanton density
$<N>/V_4\sim 1$ fm$^{-4}$.
We arrived at (\ref{1}) by integrating over a random system of instantons and
antiinstantons and then bosonizing the multi-t' Hooft vertices.
We note that eventhough $W_I$ is subleading in $1/N_c$, since the
instanton-antiinstanton density $<N>/V_4$ is of order $N_c$, a statistical
averaging of this part leads effects of order $N_c^0$ (see below), thus
comparable to the free meson action $W$.  If we were to drop this term, the
present
analysis is totally in agreement with the detailed analysis of Diakonov
and Petrov \cite{DIA1}, in the two flavour case. This term reflects on the
fermionic induced interactions and will be important for the discussion below.
We have explicitly
disregarded the scalar fields for convenience. The latters are relevant for the
issue of the scale anomaly and the bulk compressibility. We refer to
\cite{NVZ2} for a discussion of these issues. We stress that the fields $K$ are
auxillary (integrated over), while the physical meson sources are represented
by $S$. The auxillary fields arise from the bosonization of the multiflavour
quark interactions triggered by a random instanton gas. The $a$ $posteriori$
legitimacy of the random gas approximation will be discussed below.

To leading order in $N_c$ and in the long wavelength limit, the fermionic
correlation functions follow from (\ref{3}). Specifically

\be
\langle \overline\psi i\gamma_5\lambda^a\psi (x)
 \overline\psi i\gamma_5\lambda^a\psi (0) \rangle \sim &&
-N_c {\rm Tr}\left(\lambda^a S_F^+ (x)\lambda^a S_F (-x)\right)\nonumber\\&& +
\frac{<\overline{\psi}\psi >^2}{f^2} \left(
\frac{\delta^2{\rm ln}Z [S]}{\delta S^a(x)\delta S^a(0)}\right)_{S=0}
\label{31}
\ee
where $S_F (x)$ is the constituent quark propagator, and the trace
is over the spin variables.  It suffices to know, that asymptotically
$S_F(x)\sim M^2 e^{-M|x|}/|x|$, where $M\sim m+ M(0) \sim m+ 345$ MeV
\cite{DIA1,NVZ2}.
Similar expressions can be derived  in other channels. The presence of the
two-constituent quark cut in (\ref{31}) reflects on the lack of confinement
in the model. It is interesting to compare expression (\ref{31}) with
a similar expression derived in two dimensional QCD for large $N_c$
\cite{GROSS1}. In the latter, the two-fermion cut is infrared
sensitive and cancels precisely against the infrared sensitive one-gluon
exchange graph. The result is a sum of mesonic poles only.
This cancellation makes explicit use of Ward identities in
Feynman graphs. It is quantum, and thus absent in the semiclassical approach
followed here.

For finite $N_c$ the fermions feed back on the instantons and vice-versa.
A straightforward way to incorporate this, is to use (\ref{1}) for
finite $N_c$. This is an approximation, as subleading parts in the
multi-'t Hooft interactions were disregarded in (\ref{1}).
We expect these parts to
renormalize the fermion condensate and the nonet decay constant.
With this in mind and in the long wavelength limit, we can
approximate the instantons and anti-instantons by point like structures
(the effects of their sizes will be discussed below) and define the
topological charge to be ($N=N_+ +N_-$)

\be
(n^+-n^-)(x) = \sum_{i=1}^N Q_i \delta_4 (x- x_i)
\label{4}
\ee
with  $Q_{\pm} =\pm 1$. Thus, the interaction term between instantons in the
long wavelength limit is mostly triggered by the singlet $K^0$, through their
topological charge

\be
W_I = -i\frac{\sqrt{2N_F}}{f} \sum_{i=1}^N Q_i \,\,K^0 (x_i )
\label{5}
\ee
We note that the combination $(i\sqrt{f} K^0)$ plays the role of a four
dimensional Coulomb potential (the i for Euclidean),
with Coulomb charges $q_i = Q_i \sqrt{2N_F/f}$.
This point will be further discussed below.
If we denote by $z_{\pm}$ the fugacity of the quasiparticles, then
substituting (\ref{4}) in (\ref{1}) gives ($N=N_+ +N_-$)

\be
Z[ S] = \sum_{N_{\pm}} \frac {z_+^{N_+} z_-^{N_-}}{N_+! N_- !}
\prod_{i=1}^N \int d^4 x_i \int d[K]\,\,e^{-W [K] + i\frac{\sqrt{2N_F}}f
\sum_{i=1}^N Q_i \,\, K^0(x_i ) + \int d^4x \,{\rm Tr} (SK) }
\label{6}
\ee
The fugacities follows from $<N_{\pm}> = \partial\, {\rm ln}z_{\pm}\, Z [0]/
{\partial z_{\pm}}$. For a noninteracting system ${\rm ln} Z [0]\sim
V_4(z_++z_-)$. Thus $z_{\pm} \sim (<N>/2V_4)$.

In the vacuum ($S=0$) we can rewrite (\ref{6}) in two equivalent ways.
By summing over the instanton-antiinstanton degrees of freedom in the
long-wavelength limit, leaving out the singlet field. Thus

\be
Z [0] =\int d[K] \,\,e^{-W [K] + \int d^4x \,\,2z \,\,{\rm
cos}(\sqrt{2N_F}K^0/f)} =\int d[K]\,\, e^{-S[K]}
\label{7}
\ee
which shows that $Z[0]$ is a sum of zero point contributions from a free
but light octet of pseudoscalars, and a self-interacting but heavy
pseudoscalar singlet.
This version of $Z[0]$ will be relevant for the discussion of matter effects
below. Equivalently, we can choose to integrate over the singlet $K^0$ in
(\ref{6}) leaving out the instanton-antiinstanton degrees of freedom . Ignoring
singlet-octet mixing, the result is

\be
Z [0] = \left( \int d'[K] e^{-W[K]}\right)\,\,
\sum_{N_{\pm}} \frac {z^{N_++N_-}}{N_+! N_- !}
\prod_{i=1}^N \int d^4 x_i \,\,e^{-N_F \frac { m_0^2}{f^2}
\sum_{i,j} Q_iQ_j \left( \frac 1{2\pi^2}
\frac{{\bf K}_1(m_0 |x_i-x_j|)}{m_0|x_i-x_j|}\right)}
\label{8}
\ee
where $m_0^2= 2/3m_K^2 +1/3 m_{\pi}^2$ is the singlet $K^0$ mass, and ${\bf
K}_1$ a Bessel function. The singlet $K^0$ field  is excluded from the mesonic
measure, thus the prime. (\ref{8}) is the product of the vacuum
partition function of free massive octets and a four dimensional Coulomb gas
with Yukawa interactions. A similar result was also obtained by Kikuchi and
Wudka \cite{CP}. The mixing (\ref{6}) causes the
instanton-antinstantons in the vacuum to be screened over distances of the
order of $1/m_0\sim 1/2$ fm. Note that the partition function for the Coulomb
gas is ill-defined as $x_i\rightarrow x_j$. Throughout, we will assume that
the instantons and antiinstantons have a $core$ (smeared). The effects of the
smearing will cause the fugacities to renormalize by self-interaction terms.
In the long wavelength limit, the physics is insensitive to the detailed choice
of the core.

The mixing (\ref{5}) causes the $\eta'$ excitation to be heavier than the
rest of the octet. This is just the Higgs mechanism.
Indeed, the pseudoscalar correlator correlator follows from

\be
Z[\eta'] =\int d [K] e^{-W [K] +
\int d^4x \,\,2z \,\,{\rm cos}(\sqrt{2N_F}K^0/f ) + \int d^4x \eta'
\,K^0/\sqrt{2N_F} }
\label{9}
\ee
using (\ref{31}). In the Gaussian approximation (order $N_c^0$ in the action
and $N_c$ in the correlator) the result is

\be
\langle \overline\psi i\gamma_5\psi (x)
 \overline\psi i\gamma_5 \psi (0) \rangle \sim
-N_c {\rm Tr}\left(S_F^+ (x)\,\, S_F (-x)\right) +
\frac{<\overline{\psi}\psi >^2}{2N_Ff^2} \left(
\frac {m_{\eta'}^2}{2\pi^2}
\left( \frac{{\bf K}_1 (m_{\eta'} |x| )}{m_{\eta'}|x|}\right)\right)
\label{10}
\ee
where the square of the $\eta'$ mass is given by

\be
m_{\eta'}^2 = m_0^2 + 2z\,\,\frac{2N_F}{f^2}
\label{11}
\ee
which is the result quoted in \cite{NVZ2}.
For $N_F=3$, $f\sim 90 $ MeV and $z\sim 1/2$ fm$^{-4}$ we have
$m_{\eta'}\sim 1100$ MeV. The present
analysis is consistent with the analysis suggested by Witten \cite{WITTEN}
and Veneziano \cite{VENEZ},
except for one thingh : $z\sim N_c$, so that the contribution of the Coulomb
gas is $z/f^2\sim N_c^0$. The $\eta$-$\eta$' splitting does not vanish in the
large $N_c$ limit in the instanton vacuum. Should we be concerned ? Yes, if we
were to believe that the transition from $N_c=3$ to $N_c=\infty$ is smooth
in QCD (no phase change). Phase changes may occur in instanton systems with
large values of $N_c$ \cite{DIA3}. For
a dilute system $N/V_4\sim 2z\sim 1$ $ {\rm fm}^{-4}$,
the interparticle distance is about 1 fm, which is twice the screening
length $1/m_0\sim 1/2$ ${\rm fm}$. In this regime, the instantons and
antiinstantons behave as a free noninteracting gas. This result shows that
the approximations used in \cite{NVZ1,VS,DIA2,NVZ2} in which the instanton
and antiinstanton distributions were taken to be totally random is justified
in the unquenched approximation (finite $N_c$).
Finally, the induced term (\ref{6}) causes the $\eta$' to self-interact.
The $\eta$' quartic interaction term has a strength of  order
$2z (\sqrt{2 N_F}/f)^4\sim 10^2$, which is strong. We note that this
interaction vanishes in the large $N_c$ limit. Finally, we note that the
${\eta}$' pole in (\ref{10}) lies above the two-constituent quark cut,
which is about 700 MeV, leaving the $\eta'$ with a broad width. This problem is
generic to models that do not confine.

\vskip 1cm
{\bf 3. ${\rm T} = 0$  Susceptibilities}
\vskip .5cm

The topological susceptibility of the vacuum can be calculated using

\be
\chi (x-y) = \langle \sum_i Q_i \delta_4 (x-x_i) \sum_j Q_j \delta_4
(y-x_j)\rangle
\label{12}
\ee
since $E\cdot B\sim (n^+ -n^-)$. The expectation value (\ref{12}) follows from
the following generating functional

\be
Z[\theta ] =\sum_{N_{\pm}} \frac {z^{N_++N_-}}{N_+! N_- !}
\prod_{i=1}^N \int d^4 x_i \,\,e^{Q_i\theta (x_i)}\,\,e^{-N_F\frac {m_0^2}{f^2}
\sum_{i,j} Q_iQ_j
\left( \frac 1{2\pi^2}\frac{{\bf K}_1(m_0 |x_i-x_j|)}{m_0|x_i-x_j|}\right)}
\label{13}
\ee
by taking $\partial^2{\rm ln }Z/\partial\theta (x)\partial\theta (y)$ and
setting $\theta$ to zero. A straighforward calculation gives

\be
\chi (x-y) = \frac {f^2}{2N_F} (m_{\eta'}^2-m_0^2)
\left(\,\delta_4 (x-y) - (m_{\eta'}^2-m_0^2)
\frac{m_{\eta '}^2}{2\pi^2}
\left(\frac{{\bf K}_1 (m_{\eta '} |x-y |)}{m_{\eta'}|x-y|}\right)\right)
\label{14}
\ee
At large separations, the susceptibility falls off exponentially at a rate
given by the physical $\eta$' mass.  This fall off agrees with numerical
simulations \cite{VSCP}.
We note that (\ref{14}) is analogous to the result in the two dimensional
Schwinger model, where the susceptibility is given by the correlation
function of the electric field \cite{WIPF}

\be
\chi_S (x-y) =
\frac 1{4\pi^2} < E (x) E (y) > = \frac 1{4\pi^2}\left(
\delta_2 (x-y) - \frac{m_S^2}{2\pi} {\bf K}_0 (m_S |x-y|)\right)
\label{141}
\ee
with $m_S=e/\sqrt{\pi}$ the photon screening mass and ${\bf K}_0$ a Bessel
function. The topological susceptibility obeys a zero momentum
sum rule. Indeed, Fourier transforming (\ref{14})  gives at zero momentum

\be
\chi (q=0) = -\frac {2m+m_s}3 \frac {<\overline\psi \psi >}{2N_F}
\left( 1 + \frac {2m+m_s}3 \frac {<\overline\psi \psi >}{f^2m_{\eta'}^2}\right)
\label{15}
\ee
in agreement with the QCD Ward identity \cite{WARD}

\be
\int d^4 x \,\,\langle G( x) G( 0) \rangle
=- m\frac{<\overline\psi \psi >}{2N_F} +\frac{m^2}{4N_F^2}
\int d^4x \,\,\langle \overline\psi\gamma_5 \psi (x) \overline\psi\gamma_5 \psi
(0)
\rangle
\label{16}
\ee
where we have used $G= ({g^2}/{16\pi^2}) \vec{E}^a\cdot\vec{B}^a$. For the
parameters quoted above, $\chi \sim .07$ fm$^{-4}$, which is small.
Note that in the chiral limit, the topological susceptibility (\ref{15})
vanishes like in the Schwinger model where $\chi_S (q=0) =0$.

By analogy with the Schwinger model, the following expectation value

\be
\langle e^{-i\int_{D_4}\,\, d^4x G(x) }\rangle =
< \prod_i^N e^{-iQ_i\int_{D_4}d^4x \,\,\delta_4( x- x_i)}>
\label{161}
\ee
can be estimated. Here $D_4$ is a four dimensional volume with
$D_3=\partial D_4$ as a border. This expectation value reflects on the amount
of topological screening in the vacuum. Smilga \cite{SMILGA} has suggested
that (\ref{161}) obeys a four-volume law in Yang-Mills theories, and a
three-surface law in QCD. In the Schwinger model
the screening is total, and the analogue of (\ref{161}) is the Wilson
loop. The latter obeys a perimeter law ($\chi_s ( q=0) = 0$).
In our case, a semiclassical estimate of (\ref{161}) gives

\be
< \prod_i^N e^{-iQ_i\int_{D_4}d^4x \,\,\delta_4( x- x_i)}> \sim e^{-S[K^0_{\rm
cl}]}
\label{162}
\ee
where $K^0_{\rm cl}$ is the classical solution to

\be
(-\Box +m_0^2)K^0_{\rm cl} (x) +2z \frac{\sqrt{2N_F}}f {\rm sin} \left(
\frac{\sqrt{2N_F}}f K^0_{\rm cl} (x)\right) = \frac{f}{\sqrt{2N_F}}
(-\Box +m_0^2) \int_{D_4} d^4y\,\, \delta_4 (x-y)
\label{163}
\ee
for a strongly coupled source. While we do not know of a general solution to
(\ref{163}) in four dimensions, in the linear approximation (\ref{163}) can be
readily solved. In this case, the solution is equivalent to taking the
second cumulant in (\ref{161}). The latter is just the topological
susceptibility,

\be
< \prod_i^N e^{-iQ_i\int_{D_4}d^4x\,\, \delta_4( x- x_i)}> \sim e^{-\frac
{D_4}2
\chi (q=0)}
\label{164}
\ee
Since (\ref{15}) does not vanish, the second cumulant
indicates that (\ref{161}) obeys a four-volume law. The fall off rate
is $\chi /2\sim 10^{-2}$ fm$^{-4}$. This result is not exact
and maybe affected by higher cumulants (non-Gaussian fluctuations). We note
that the cumulant expansion is consistent with the dilute gas approximation.
In the chiral limit, the topological susceptibility vanishes, and the screening
is total. Indeed, for $m=m_s=0$ (\ref{164}) falls off with the border
$D_3=\partial D_4$

\be
< \prod_i^N e^{-iQ_i\int_{D_4}d^4x \,\,\delta_4( x- x_i)}>\sim
e^{-\frac{1}2 (\frac{zm_*^2}{\pi^2})\int_{D_3\times D_3}\,d\vec\Sigma_x\cdot
d\vec\Sigma_y \,\left(\frac{{\bf K}_1(m_* |x-y|)}{m_* |x-y|}\right)} =
e^{-(\frac{2z}{m_*}) D_3}
\label{165}
\ee
where $m_*^2=4N_Fz/f^2$ with $d\vec\Sigma$ the three-normal to the four-volume
$D_4$. The fall-off rate is given by

\be
\frac {2z}{m_*} = \frac {f^2}{2N_F} \left(m_{\eta'}^2 -\frac 23 m_K^2 -\frac 13
m_{\pi}^2\right)^{1/2}
\label{166}
\ee
which is about $1$ fm$^{-3}$.
(\ref{165}) follows from the second cumulant and large three-surfaces
$D_3m_*^3 >>1$. This result is analogous to the result in the Schwinger model.
In the chiral limit, the instantons and antinstantons
are in a strongly coupled plasma-like phase driven by a Debye-Huckel
equation, with an effective temperature $T_*\sim f$ (see below).

The fluctuations in the number density of instantons and antiinstantons
can be obtained using similar arguments, if we were to note that

\be
n^+ + n^- = \sum_{i}^N \delta_4 (x-x_i)
\label{167}
\ee
Since we have ignored the scalars  we will only
provide a simple estimate in this case. A straighforward calculation gives

\be
\sigma (x-y) \sim \langle \sum_{i=1}^N \delta_4 (x-x_i)
\sum_{i=j}^N \delta_4 (y-x_j) \rangle_C \sim 2z \delta (x-y)
\label{168}
\ee
where only the connected part has been retained.  This result
shows that the variance in the number of particles is Poissonian,
and that the compressibility is of order $1$ fm$^{-2}$.
The inclusion of the scalars  smears the correlations (\ref{168}) over the
scalar Compton wavelength, causing the compressibility to increase.
These effects along with the QCD scale
Ward identities \cite{SHIF} will be discussed elsewhere \cite{NZ}.

The present interplay between the instantons and the $\eta$' with the
subsequent screening, does not affect the pseudoscalar octet in the long
wavelength approximation. Indeed, from (\ref{1}) it follows that the auxillary
meson fields decouple from the topological charge to order $N_c^0$,
to the exception of the
singlet $K^0$ (ignoring singlet-octet mixing). A self-consistent calculation
would require that instead of using a random gas approximation, we should
use a screened Coulomb gas approximation, $e.g.$ (\ref{13}).
However, the diluteness of the
system implies that the effects are small, since the interparticle distance
is about twice the screening length. Thus we would predict that
the fermion condensate and the nonet decay constant will not be affected
considerably by the screening mechanism. In a way, this is good because it
implies that most of the calculations performed in the random gas approximation
reflect correctly on unquenched QCD \cite{VS,DIA2,NVZ2}.

\vskip 1cm
{\bf 4. ${\rm T} \neq 0$ Susceptibilities}
\vskip .5cm

How does the temperature affects the present arguments ? First, consider the
correlation in the topological charge (\ref{12}). At low temperature, the
system is meson dominated. The temperature enters both through the meson
parameters \footnote{The fermion constituent mass becomes temperature dependent
\cite{DIA4,NVZ3}.} as well as the meson distributions. Thus ($\omega_n =2\pi
nT$)

\be
\chi (x, y ;T) = &&+\frac{f^2(T)}{2N_F} (m_{\eta'}^2(T) -m_0^2(T))
\delta_4(x-y)\\&&
 -\frac{Tf^2(T)}{4N_F} \frac{(m_{\eta'}^2(T) -m_0^2(T))^2}{|\vec x-\vec y|}
\sum_{n=-\infty}^{+\infty} e^{i\omega_n (x^0-y^0) -\sqrt{\omega_n^2
+m^2_{\eta'}(T)} |\vec x-\vec y |}\nonumber
\label{170}
\ee
The static part $\chi (\omega =0, \vec x ; T)$ of (27)
is dominated by the $\eta'$ excitation

\be
\chi (\omega =0, \vec x; T) = && T \int_0^{\frac 1T} dx^0 \chi (x, 0; T)
\nonumber\\= &&
\frac{Tf^2(T)}{2N_F} (m_{\eta'}^2(T) -m_0^2(T)) \left(
\delta_3 ( x ) -
\frac{(m_{\eta'}^2(T) -m_0^2(T))}{2|\vec x |}
\,\,e^{-m_{\eta'}(T) |\vec x |}\right)\nonumber\\
\label{171}
\ee
In this form, the QCD Ward identity is fulfilled even at finite
temperature,

\be
\int d^3 x \chi (\omega=0, \vec x , T) =
-\frac{2m+m_s}3 \frac{<\overline\psi \psi>(T)}{2N_F}
\label{172}
\ee
and vanishes at the chiral transition point. This, however, does not
necessarily mean that the $U_A(1)$ symmetry is restored. The restoration of
the latter can be asserted only through the large distance behaviour of
(\ref{171}). In the $U_A(1)$ broken phase, the
correlator falls off exponentially, while in the symmetric phase it vanishes
identically. This point is worth probing both on the lattice and in
the numerical
instanton calculations \cite{VS}. For completeness, we note that
the temporal asymptotics of (\ref{170})
is free field dominated for temperatures $T\sim m_{\eta '}/2\pi \sim 150$ MeV
($0<x^0<1/T$). The compressibility $\sigma (T) \sim 2z (T)$, and is
expected to drop by $1/2$ at the chiral transition point, following the drop
in the gluon condensate \cite{ADAMI}. Debye screening at high temperature
\cite{GROSS2} causes the electric condensate to vanish. In instanton models
that would also mean the vanishing of the magnetic condensate because of self
duality.

In the Schwinger model, the temperature effects on the topological correlator
can be calculated  exactly. Indeed, the correlation in the topological
charge at finite temperature is given by

\be
\chi_S (x, T) = \frac 1{4\pi^2} \left(
\delta_2 (x) - \frac {Tm_S^2}{2\pi}\sum_{n=-\infty}^{+\infty}
\frac {1}{\sqrt{\omega_n^2 +m_S^2}}e^{i\omega_n x^0}
e^{-\sqrt{\omega_n^2+m_s^2}
|x^1|} \right)
\label{174}
\ee
which is the analogue of (27).
The analogue of the static part of the susceptibility (\ref{171}) is given by

\be
\chi_S (\omega=0, x^1 ;T) =\int_0^{\frac 1T}\chi_S (x, T) = \frac T{4\pi^2}
\left( \delta (x^1) -\frac {m_S}2 e^{-m_S|x^1|}\right)
\label{175}
\ee
whatever $T$.  We observe that
eventough (\ref{175}) integrates to zero, the correlator still falls
off exponentially. In the Schwinger model $m_S$ follows directly from the
$U_A(1)$
anomaly (bubble graph) and thus is $T$-independent for all temperatures. This
is not the case for $m_{\eta'}$ as we have discussed above.
For completeness, we also note that along the temporal
direction (\ref{174}) is dominated by the free field behaviour for temperatures
of the order of $T\sim m_S/2\pi$ ($0<x^0<1/T$),

\be
\chi_S(x^0, T) \sim \frac 1{4\pi^2}\left(
V_1\delta (x^0) - \frac {Tm_S}{2\pi} + \frac{m_S^2}{2\pi^2}\,\,
{\rm ln} (2\pi T |x^0|)\right)
\label{176}
\ee
where $V_1$ is the space length.

\vskip 1cm
{\bf 5. Pressure}
\vskip .5cm

At low temperature, the pressure is just given by the one loop
effects following from (\ref{1}) and the rest
of the mesonic action. Thus

\be
{\bf P} + {\bf B}= {\bf P}_{\pi} + {\bf P}_{K} + {\bf P}_{\eta} + {\bf
P}_{\eta'} + ...
\label{17}
\ee
where ${\bf B}$ is the vacuum pressure, and ${\bf P}_a$ the respective
mesonic pressures.
The dots stand for the higher hadron resonances, in agreement with the Gibbs
average. The mesons in (\ref{17}) carry temperature dependent
masses, since the fermion condensate is temperature dependent. At low
temperature, the effects are small, and the pressures ${\bf P}_a$ in
(\ref{17}) are the usual black-body contributions.
Below the deconfining transition, the vacuum pressure
is $T$-independent in a confining theory \cite{HAN}, to leading order in $N_c$.
This is not the case here and ${\bf B}$ receives an unwanted contribution from
the free constituent quark loop of order $N_c$.

In writing (\ref{17}) we have ignored the $\eta$'
self-interactions, $i.e.$ we have kept the order $N_c^0$ terms in the free
energy. For finite $N_c$ these effects, may not be small. Indeed,  strong
self-interactions may give rise to classical solutions to
($\phi = i \sqrt{f} K^0$)

\be
(-\Box + m_0^2 ) \,\,\phi = -z \left(\frac{2N_F}{f}\right)^{1/2}
\left( e^{-\frac{\sqrt{2N_F/f}}f \phi} -e^{+\frac{\sqrt{2N_F/f}}f \phi}\right)
\label{18}
\ee
which shows that $\phi$ (the precursor of the $\eta'$)
plays the role of a coarse grained Coulomb
potential in four dimensions. In the chiral limit
($m_0=0$) (\ref{18}) is the Debye-Huckel equation for
a four dimensional Coulomb plasma with charges
$q=\sqrt{2N_F/f}\sim 1/\sqrt{N_c}$, charge
density $n\sim zq/{4\pi }\sim \sqrt{N_c}$, and an effective temperature
$T_*\sim f\sim \sqrt{N_c}$
\footnote{A four dimensional plasma, follows from a five dimensional field
theory with $\phi\sim M^{3/2}$ and $q\sim{M}^{-1/2}$ where
$M$ carries canonical mass dimension.}.
For large $N_c$, the Coulomb plasma is classical, and (\ref{17}) follows.
The $\eta-\eta'$ mass difference is just the Debye screening length.
At high temperature, the effects of the dropping
masses may be important. The derivation, however, becomes less reliable as
other
effects (quantum fluctuations, subleading $1/N_c$ terms, ...) may be important.
Note that at high temperature the four dimensional
Debye-Huckel equation (\ref{18}) (taken litterally)
dimensionally reduces to an equation in
three dimensions. The effective temperature
for the three dimensional Coulomb gas is $f\sim 0$, that is low. The system
has a tendency to cluster. This tendency is perhaps what has been observed
by Ilgenfritz and Shuryak \cite{IIL} in interacting instanton systems with
fermion determinants. In a way the high temperature Coulomb gas problem
discussed here behaves as a low temperature Kosterlitz-Thouless system
\cite{KT}. Coulomb systems in two dimensions display a dipole phase at low
temperature, and a plasma phase at high temperature.
If such a transition were to occur, it is plausible that the resulting phase
is $U_A(1)$ symmetric.
Since $\chi (0)\sim m <\overline\psi \psi >$, it is possible that this phase
change
may coincide with the chiral transition. This point can be elucidated by
lattice simulations that will calculate both $\chi (\omega =0, \vec q =0, T)$
and $\chi (\omega = 0, |\vec x |, T)$, as suggested above.

We note that the effects of the instantons in (\ref{17}) is implicit, and
manifest through the larger $\eta$' mass even at finite temperature
(ignoring $\eta$' interactions). An equivalent way of describing the
same physics, is to rewrite (\ref{17}) as follows

\be
{\bf P}+{\bf B} = {\bf P}_{\pi} + {\bf P}_{K} + {\bf P}_{\eta} +
\left( {\bf P}_{K^0} + P_I \right) + ...
\label{19}
\ee
where $P_{K^0}$ is the contribution to the pressure coming from a thermalized
system of singlet mesons with mass $m_0$, and $P_I$ the instanton contribution
following from the finite temperature partition function

\be
Z_I[ T] = \sum_{N_{\pm}} \frac {z^{N_++N_-}}{N_+! N_- !}
\prod_{i=1}^N \int_{R^3\times 1/T} d^4 x_i \,\,
e^{-N_F \frac {T}{ f^2 ( T)}
\sum_{i,j,n} Q_iQ_j \int \frac{d^3q}{(2\pi)^3}
\frac {e^{iq\cdot (x_i-x_j)}}{q_n^2 + m_0^2(T )}}
\label{20}
\ee
where $q_n =(2\pi n T, \vec q )$. $P_{\eta '}= P_{K^0} + P_I$.
The Coulomb gas description referred to above is now manifest.
This decomposition, allows us to see  how matter
affects instantons. It also shows the artificial character of the decomposition
in (\ref{19}). At low temperature, the condensate
(thus $m_0$) and the meson decay constant $f$ do not change appreciably. The
instanton screening length remains about the same. At high temperature $m_0$
is expected to drop, weakening the screening. Our calculations break
down when the screening becomes comparable to the interparticle distance,
$i.e.$ $m_0^4\sim <N>T/V_3$. Subleading terms in $1/N_c$ and the omitted
scalars
as well as higher lying hadrons are important at high temperature.

\vskip 1cm
{\bf 6. Conclusions}
\vskip .5cm

We have given some qualitative arguments indicating the
interplay between the $\eta$' fluctuations and the instantons in the vacuum
state and in matter. At zero temperature,
the instantons are screened over distances of the order of
$1/2$ fm, and the $\eta$' acquires a mass of the order of 1100 MeV, leading to
a natural resolution of the $U_A(1)$ problem, as originally suggested by
't Hooft \cite{HOOFT}. These results emphasize our earlier work \cite{NVZ2}
and are in general agreement with the arguments given by Kikuchi and Wudka
\cite{CP}. The screening  among the
instantons and antiinstantons in the vacuum yields a topological
susceptibility that is consistent with the QCD Ward identity. For dilute
systems of instantons, the screening justifies the use of the random gas
approximation for finite $N_c$ (unquenched calculations)
\cite{VS,DIA2,NVZ2}. In the instanton
model of the vacuum the $\eta-\eta'$ mass splitting does not vanish in the
large $N_c$ limit.

The screening is unaffected by
low temperatures, $i.e.$ temperatures for which the pion and kaon mass do not
change substantially. As expected, the bulk pressure is meson dominated.
The $\eta$' contribution to the pressure can be viewed as originating
either from a strongly correlated and massive $\eta$' at temperature $T$,
or from a free light singlet at temperature $T$ mixed to
a four dimensional Coulomb gas  with an effective temperature $f(T)$.
The latter can be described by a Debye-Huckel equation in four dimensions.
At high temperature substantial changes in the bulk
parameters together with strong hadronic correlations make a simple
assessment of the instanton contribution to the pressure difficult.
The screening, however, is found to decrease at high temperature.
Our estimates in matter becomes unreliable at temperatures of the order of
$T \sim m_0^4 <N>/V_3$. At these temperatures, a Kosterlitz-Thouless transition
in the instanton language may occur, followed by a substantial drop in
the topological susceptibility, since $\chi (q=0)\sim m<\overline\psi \psi>\sim
0$.
This transition is likely to be $U_A(1)$ restoring and may be related to
the chiral restoration transition in QCD. This point can be clarified by
studying the static  correlations in the topological charge at large spatial
separations.

In so far, our attitude has been to constrain the instanton effects by working
from the long
wavelength limit, where the physics is known.
Alternatively, one can try to
start from the short wavelength limit, and build up the correlations from
the original fermion determinant in the QCD action \cite{VS}.  We believe
that our arguments provide helpful constraints in the long wavelength limit
both in the vacuum and in matter.

\vskip 1.5cm
{\bf \noindent  Acknowledgements \hfil}
\vglue .5cm

I would like to thank Maciek Nowak and Jac Verbaarschot for discussions.
This work was supported in part by the US DOE grant DE-FG-88ER40388.

\vfill
\eject
\setlength{\baselineskip}{15pt}


\end{document}